\documentclass[12pt]{article}
\newcommand{\bq}{\begin{equation}}
\newcommand{\eq}{\end{equation}}
\newcommand{\ba}{\begin{eqnarray}}
\newcommand{\ea}{\end{eqnarray}}

\newcommand{\nl }{ \nonumber  }

\newcommand{\p}{\partial}
\newcommand{\h}{\hspace{.5cm}}
\newcommand{\s}{\sigma}

\newcommand{\la}{\lambda}
\newcommand{\La}{\Lambda}
\newcommand{\Di}{\left(\p_0-\la^{i}\p_i\right)}
\newcommand{\Dj}{\left(\p_0-\la^{j}\p_j\right)}

\begin{document}
\begin{center}
{\bf MEMBRANE SOLUTIONS IN M-THEORY 
\vspace*{0.5cm}\\ P. Bozhilov}
\\ {\it Institute for Nuclear Research and Nuclear Energy, \\
Bulgarian Academy of Sciences, \\ 1784 Sofia, Bulgaria\\
E-mail:} bozhilov@inrne.bas.bg
\end{center}
\vspace*{0.5cm}

Motivated by the recent achievements in the framework of the semiclassical limit of the M-theory/field theory correspondence, we propose an approach for obtaining exact membrane solutions in general enough M-theory backgrounds, having field theory dual description. As an application of the derived general results, we obtain several types of membrane solutions in $AdS_4\times S^7$ M-theory background.


\vspace*{.5cm} 


\section{Introduction}
In the last few years, a remarkable progress has been achieved in understanding the semiclassical limit of the string/gauge theory duality, where the classical string solutions in curved backgrounds play an important role. Expectedly, attempts have been made to extend this success to M-theory/field theory correspondence and in particular, to obtain new membrane solutions and relate their conserved charges to the dual objects on the field theory side \cite{6}-\cite{BRR04}. 

M2-brane configurations in $AdS_7\times S^4$ space-time, with field theory dual $A_{N-1}(2,0)$ $SCFT$, have been considered in \cite{6}-\cite{14}. In \cite{6}, rotating membrane solution in $AdS_7$ have been obtained. Rotating and boosted membrane configurations was investigated in \cite{10}. Multiwrapped circular membrane, pulsating in the radial direction of $AdS_7$, has been considered in \cite{14}. The article \cite{27} is devoted mainly to the investigation of rotating membranes on $G_2$ manifolds. However, membrane configurations in $AdS_4\times S^7$, $AdS_4\times Q^{1,1,1}$, warped $AdS_5\times M^6$ and in 11-dimensional $AdS$-black hole backgrounds have been also considered. In \cite{HT04} and \cite{BRR04}, new membrane solutions in $AdS_p\times S^q$ have been also obtained, by using different type of membrane embedding.

To further develop the above achievements, it will be useful to have a method for obtaining exact membrane solutions in general enough M-theory backgrounds, having gauge theory dual description.

In \cite{NPB656}, an approach has been proposed, which give the possibility for obtaining explicit exact solutions of the equations of motion and constraints for $p$-brane and $Dp$-brane embedding, such that the resulting dynamics depends only on the worldvolume time parameter $\tau$. In \cite{JHEP032}, this method has been applied to membranes ($p$=2) in $AdS_7\times S^4$ background. Here, in section 2, we extend the above approach to membrane embedding, which lead to $\s$-dependent dynamics, where $\s$ is one of the spatial vorldvolume coordinates. In section 3, as an application of the derived general results, we consider membranes moving in $AdS_4\times S^7$ space-time.

\setcounter{equation}{0}
\section{Membrane dynamics in generic background}
In what follows, we will use an action for membrane moving in 
curved space-time with metric tensor $g_{MN}(x)$, and interacting with a 
background 3-form gauge field $b_{MNP}(x)$, which is given by
\ba\label{oma} S= \int d^{3}\xi\mathcal{L} = 
\int d^{3}\xi\left\{\frac{1}{4\lambda^0}\Bigl[G_{00}-2\lambda^{j}G_{0j}+\lambda^{i}
\lambda^{j}G_{ij}-\left(2\lambda^0T_2\right)^2\det G_{ij}\Bigr] + 
T_2 B_{012}\right\},\ea 
where \ba\nl G_{mn}= g_{MN}(X)\p_m X^M\p_n X^N,\h
B_{012}= b_{MNP}(X)\p_{0}X^{M}\p_{1}X^{N}\p_{2}X^{P}, \\ \nl 
\p_m=\p/\p\xi^m,\h m = (0,i) = (0,1,2),\h M = (0,1,\ldots,10),\ea 
are the fields induced on the membrane worldvolume,
$\lambda^m$ are Lagrange multipliers, $x^M=X^M(\xi)$ are the membrane
embedding coordinates, and $T_2$ is its tension. 
As shown in \cite{NPB656}, the above action is
classically equivalent to the Nambu-Goto type action 
\ba\label{NGa} S^{NG}= - T_2\int d^{3}\xi
\left(\sqrt{-\det{G_{mn}}}-\frac{1}{6}\varepsilon^{mnp}
\p_{m}X^{M}\p_n X^N \p_{p}X^{P} b_{MNP}\right)\ea
and to the Polyakov type action
\ba\label{Pa} S^{P}= - \frac{T_2}{2}\int
d^{3}\xi\left[\sqrt{-\gamma}\left(\gamma^{mn} G_{mn}-1\right) - 
\frac{1}{3} \varepsilon^{mnp}\p_{m}X^{M}\p_nX^N\p_{p}X^{P}b_{MNP}\right],\ea
where $\gamma^{mn}$ is the auxiliary worldvolume metric and $\gamma=\det\gamma_{mn}$.

We choose to use the action (\ref{oma}), because it possesses the
following advantages. First of all, it does not contain square root 
(as the  Nambu-Goto type action), thus avoiding the introduction of additional 
nonlinearities in the equations of motion. Besides, the equations of motion 
for the Lagrange multipliers $\lambda^m$ generate the {\it independent} 
constraints only (opposite to the Polyakov type action). Finally, this 
action gives a unified description for the tensile and tensionless membranes, 
so the limit $T_2\to 0$ may be taken at any time 
(opposite to both other actions).

Further on, we will work in the worldvolume gauge $\lambda^m=constants$, 
in which the equations of motion for $X^M$, following from (\ref{oma}), 
are given by $(\mathbf{G}\equiv\det{G_{ij}})$
\ba\label{eqm} &&g_{LN}\left[\Di\Dj X^N - \left(2\lambda^0T_2\right)^2
\p_i\left(\mathbf{G}G^{ij}\p_j X^N\right)\right]\\ \nl
&&+\Gamma_{L,MN}\left[\Di X^M \Dj X^N - \left(2\lambda^0T_2\right)^2
\mathbf{G}G^{ij}\p_i X^M \p_j X^N\right]\\ \nl
&&=2\la^0 T_2 H_{LMNP}\p_0X^{M}\p_1X^N\p_2X^{P},\ea
where
\ba\nl \Gamma_{L,MN}=g_{LK}\Gamma^K_{MN}=\frac{1}{2}\left(\p_Mg_{NL}
+\p_Ng_{ML}-\p_Lg_{MN}\right)\ea
are the components of the symmetric connection corresponding to the metric
$g_{MN}$ and $H_{LMNP}$ is the field strength of the
$3$-form gauge field $b_{MNP}$.

By varying the action (\ref{oma}) with respect to $\lambda^m$, one obtains 
the constraints 
\ba\label{0} &&4\lambda^0\left(T_0^0 - \lambda^i T_i^0\right)=
G_{00}-2\lambda^{j}G_{0j}+\lambda^{i}\lambda^{j}G_{ij}
+\left(2\lambda^0T_2\right)^2\mathbf{G}=0,\\
\label{0j} &&2\lambda^0 T_j^0=
G_{0j}-\lambda^{i}G_{ij}=0.\ea
To simplify the constraint (\ref{0}), we put (\ref{0j}) in it, 
which results in
\ba\label{00} 4\lambda^0 T_0^0=G_{00}-\lambda^{i}\lambda^{j}G_{ij}
+\left(2\lambda^0T_2\right)^2\mathbf{G}=0.\ea
Thus, our {\it independent} constraints, with which we will work
from now on, are given by (\ref{00}) and (\ref{0j}), where $T_0^0$ and $T_i^0$ are the independent components of the worldvolume stress-energy tensor. In practice, it turns out that using the diagonal worldvolume gauge $\lambda^{i}=0$ simplify the considerations a lot, as we will see later on\footnote{This corresponds to the gauge choice in \cite{27}, where the Polyakov type action (\ref{Pa}) was used.}. In this case, the constraints reduce to 
\ba\nl &&G_{00}+\left(2\lambda^0T_2\right)^2\mathbf{G}=0,\\ \label{rci} &&G_{0i}=0.\ea 
In particular, this means that the induced metric must be block-diagonal. 

Here, we will investigate the membrane dynamics in the framework of
the following type of embedding $(\Lambda^\mu_m = constants)$
\ba\label{sLA} X^\mu(\xi^m)=X^\mu(\tau, \delta, \sigma)=\Lambda^\mu_m \xi^m=
\Lambda^\mu_0\tau+\Lambda^\mu_1\delta+ \Lambda^\mu_2\s,\h
X^a(\xi^m)=Z^a(\s).\ea 
In the above formulas, the embedding coordinates $X^M(\xi^m)$ are divided into
$X^M=(X^\mu,X^a)$, where $X^\mu(\xi^m)$ correspond to the space-time
coordinates $x^\mu$, on which the background fields do not depend
\ba\label{ob} \p_\mu g_{MN} =0,\h \p_\mu b_{MNP} =0.\ea
In other words, we suppose that there exist $n_\mu$ commuting Killing vector fields
$\p/\p x^\mu$, where $n_\mu$ is the number of the coordinates $x^\mu$.

Taking into account the ansatz (\ref{sLA}), one obtains that the fields induced on 
the membrane worldvolume are given by (the prime is used for $d/d\sigma$)
\ba\nl &&G_{00}=\Lambda^\mu_0\Lambda^\nu_0 g_{\mu\nu},\h
G_{11}=\Lambda^\mu_1\Lambda^\nu_1 g_{\mu\nu}, \h G_{22}=g_{ab}Z'^a Z'^b +
2\Lambda^\mu_2 g_{\mu a}Z'^a + \Lambda^\mu_2\Lambda^\nu_2 g_{\mu\nu},\\ \nl 
&&G_{01}=\Lambda^\mu_0\Lambda^\nu_1 g_{\mu\nu},\h 
G_{02}=\Lambda^\mu_0\left(g_{\mu a}Z'^a + \Lambda^\nu_2 g_{\mu\nu}\right),\h
G_{12}=\Lambda^\mu_1\left(g_{\mu a}Z'^a + \Lambda^\nu_2 g_{\mu\nu}\right),\\ \label{if}
&&B_{012}=\Lambda^\mu_0\Lambda^\nu_1\left(b_{\mu\nu a}Z'^a  + 
\Lambda^\rho_2 b_{\mu\nu\rho}\right).\ea
Correspondingly, the Lagrangian density in the action (\ref{oma}) reduces to 
\ba\label{old} \mathcal{L}^{A}(\sigma) =\frac{1}{4\lambda^0}\left[K_{ab}(g)Z'^aZ'^b + 
2A_{a}(g,b)Z'^a - V(g,b)\right],\ea where 
\ba\nl K_{ab}(g)&=&(\lambda^{2})^2g_{ab}-\left(2\lambda^0T_2\right)^2 \Lambda^\mu_1\Lambda^\nu_1
\left.\right.\left(g_{ab}g_{\mu\nu}-g_{a\mu}g_{b\nu}\right),\\ \nl 
A_{a}(g,b)&=&-\lambda^2\left(\Lambda^\mu_0-\lambda^i\Lambda^\mu_i\right)g_{a\mu}
+\left(2\lambda^0T_2\right)^2 \Lambda^\mu_1\Lambda^\nu_1\Lambda^\rho_2
\left(g_{a\mu}g_{\nu\rho}-g_{a\rho}g_{\mu\nu}\right)+\\ \nl
&&2\lambda^0T_2\Lambda^\mu_0\Lambda^\nu_1 b_{a\mu\nu},\\ \nl
V(g,b)&=&-\left(\Lambda^\mu_0-\lambda^i\Lambda^\mu_i\right)\left(\Lambda^\nu_0-\lambda^j\Lambda^\nu_j\right) g_{\mu\nu}+ \left(2\lambda^0T_2\right)^2 \Lambda^\mu_1\Lambda^\nu_1\Lambda^\rho_2\Lambda^\lambda_2
\left(g_{\mu\nu}g_{\rho\lambda}-g_{\mu\rho}g_{\nu\lambda}\right)\\ \nl
&&-4\lambda^0T_2\Lambda^\mu_0\Lambda^\nu_1\Lambda^\rho_2 b_{\mu\nu\rho}.\ea
$\mathcal{L}^{A}$ does not depend on $\tau$ and $\delta$ because of (\ref{sLA}) and (\ref{ob}).

Let us introduce the densities
\ba\nl P_M=\frac{\p\mathcal{L}}{\p\left(\p_0 X^M\right)},\h 
\mathcal{P}^i_M=\frac{\p\mathcal{L}}{\p\left(\p_i X^M\right)}.\ea
Then, in view of (\ref{sLA}) and (\ref{ob}), $P_M$ and $\mathcal{P}^i_M$ do not depend on $\tau$ and $\delta$, 
and the equations of motion (\ref{eqm}) acquire the form 
\ba\label{ecm} \left[\mathcal{P}^2_\mu(\s)\right]'=0,
\\ \label{rem} \left(\mathcal{P}^2_a\right)' - \frac{\p\mathcal{L}^A}{\p Z^a}=0.\ea
The equations (\ref{ecm}) just state that $\mathcal{P}^2_\mu$ are constants of the motion:
\ba\nl &&2\lambda^0\mathcal{P}^2_\mu = \left[(\lambda^{2})^2g_{\mu a} + \left(2\lambda^0T_2\right)^2 \Lambda^\nu_1\Lambda^\rho_1\left(g_{\mu\nu}g_{\rho a}-g_{\nu\rho}g_{\mu a}\right)\right]Z'^a
-\lambda^{2}\left(\Lambda^\nu_0-\lambda^j\Lambda^\nu_j\right)g_{\mu\nu}
\\ \label{cm} &&+\left(2\lambda^0T_2\right)^2 \Lambda^\nu_1\Lambda^\rho_1\Lambda^\lambda_2
\left(g_{\mu\nu}g_{\rho\lambda}-g_{\mu\lambda}g_{\nu\rho}\right) 
+ 2\lambda^0T_2\Lambda^\nu_0\Lambda^\rho_1 b_{\mu\nu\rho}=constants.\ea

On the other hand, as far as $\mathcal{L}^{A}$ does not depend on $X^\mu$, the momenta 
\ba\label{cmom} p_\mu &=& \int d^{2}\xi P_\mu 
\\ \nl &=& \frac{1}{2\lambda^{0}}\int\int d\delta d\sigma\left[\left(-\lambda^2 g_{\mu a} 
+ 2\lambda^0 T_2\Lambda^\nu_1 b_{\mu\nu a}\right)Z'^a + \left(\Lambda^\nu_0-\lambda^j\Lambda^\nu_j\right)g_{\mu\nu}
+ 2\lambda^0T_2\Lambda^\nu_1\Lambda^\rho_2 b_{\mu\nu\rho}\right]\ea
are conserved, i.e. they do not depend on the proper time $\tau$.

The remaining equations (\ref{rem}) may be rewritten as
\ba\label{ema} K_{ab}Z''^b + \Gamma^{K}_{a,bc}Z'^b Z'^c - 2\p_{[a}A_{b]}Z'^b  
+ \frac{1}{2}\p_a U = 0,\ea where 
\ba\nl &&\Gamma^{K}_{a,bc}= \frac{1}{2}\left(\p_b K_{ca}+\p_c K_{ba}-\p_a K_{bc}\right),
\\ \nl &&\p_{[a}A_{b]}=\frac{1}{2}\left(\p_a A_b - \p_b A_a\right), \h U=V + 4\lambda^0\Lambda^\mu_2\mathcal{P}^2_\mu.\ea

The constraints (\ref{00}) and (\ref{0j}) take the form
\ba\label{00e} &&K_{ab}Z'^aZ'^b + U=0,
\\ \label{01} &&\Lambda^\mu_1\left[\lambda^2g_{\mu a}Z'^a - \left(\Lambda^\nu_0-\lambda^j\Lambda^\nu_j\right) g_{\mu\nu}\right]=0,
\\ \label{02} &&\lambda^2g_{ab}Z'^aZ'^b + \left[\lambda^2\Lambda^\mu_2 - \left(\Lambda^\mu_0-\lambda^i\Lambda^\mu_i\right)\right]g_{\mu a}Z'^a + 
\Lambda^\mu_2\left(\Lambda^\nu_0-\lambda^j\Lambda^\nu_j\right) g_{\mu\nu}=0.\ea

Thus far, based on (\ref{sLA}) and (\ref{ob}), we reduced the membrane dynamics to the problem of solving the equations of motion (\ref{ema}), and constraints (\ref{00e}), (\ref{01}) and (\ref{02}).
Let us first consider the last two constraints. If we choose the worldvolume gauge $\lambda^2=0$, they reduce to
\ba\nl \left(\Lambda^\mu_0-\lambda^1\Lambda^\mu_1\right)\Lambda^\nu_1 g_{\mu\nu}=0,
\h\left(\Lambda^\mu_0-\lambda^1\Lambda^\mu_1\right)\left(g_{\mu a}Z'^a - 
\Lambda^\nu_2 g_{\mu\nu}\right)=0,\ea
and are obviously satisfied identically for $\Lambda^\mu_0=\lambda^1\Lambda^\mu_1$. Another type of solution may be obtained in diagonal worldvolume gauge $\lambda^i=0$, when the background metric is block-diagonal, i.e. $g_{\mu a}=0$, which is often the case. Then, (\ref{01}) and (\ref{02}) take the form
\ba\nl \Lambda^\mu_0\Lambda^\nu_1 g_{\mu\nu}=0,
\h\Lambda^\mu_0\Lambda^\nu_2 g_{\mu\nu}=0.\ea
These equalities can be satisfied identically with appropriate choice of the embedding parameters $\Lambda^\mu_m$.
In both cases, our next task is to consider the remaining equations (\ref{ema}) and (\ref{00e}). They have the same form as for the different type of membrane embedding considered in \cite{NPB656,JHEP032}, and therefore, may be treated analogously. In our notations, the results are the following.

If the embedding is such that the background seen by the membrane depends on only one coordinate $x^a$ , then the constraint (\ref{00e}) is first integral of the equation of motion (\ref{ema}) for $X^a(\xi^m)=Z^a(\s)$, and the general solution is given by
\ba\label{1dc}\sigma\left(X^a\right)=\sigma_0 + \int_{X_0^a}^{X^a}
\left(-\frac{U}{K_{aa}}\right)^{-1/2}dx,\ea 
where $\sigma_0$ and $X_0^a$ are arbitrary constants.

When the background felt by the membrane depends on more than one coordinate
$x^a$, the first integrals of the equations of motion for
$Z^a(\s)=(Z^r, Z^\alpha)$, which also solve the constraint
(\ref{00e}), are \ba\label{fir} &&\left(K_{rr}Z'^r\right)^2 = K_{rr}
\left[(n_\alpha-1)U -2n_\alpha\left(A_r-\p_r
f\right)Z'^r -\sum_{\alpha}\frac{D_{\alpha}
\left(Z^{a\ne\alpha}\right)} {K_{\alpha\alpha}}\right],\\ \label{fia}
&&\left(K_{\alpha\alpha}Z'^\alpha\right)^2 =D_{\alpha}
\left(Z^{a\ne\alpha}\right) + K_{\alpha\alpha}\left[
2\left(A_r-\p_r f\right)Z'^r-U\right],\ea where $Z^r$ is one of the
coordinates $Z^a$, $Z^\alpha$ are the remaining ones, $n_\alpha$
is the number of $Z^\alpha$, and $D_{\alpha}$ are
arbitrary functions of their arguments. The above expressions are
valid, if the "metric" $K_{ab}$ is diagonal one, and the
following integrability conditions hold
\ba\nl &&A_a \equiv(A_r,A_\alpha)=
(A_r,\p_\alpha f),\h
\p_\alpha\left(\frac{K_{\alpha\alpha}}{K_{aa}}\right)=0,
\\ \nl &&\p_\alpha\left(K_{rr}Z'^r\right)^2=0,\h
\p_r\left(K_{\alpha\alpha}Z'^\alpha\right)^2=0. \ea

\setcounter{equation}{0}
\section{An example}
In this section, as an example, we will apply our approach to a membrane moving in $AdS_4\times S^7$ M-theory background. The metric and the three-form gauge field are given by
\ba\nl &&ds^2_{AdS_{4}\times S^7} = l_{11}^2\left[-\cosh^2\rho dt^2 + d\rho^2 + \sinh^2\rho
\left(d\alpha^2 + \sin^2\alpha d\beta^2\right)+B^2ds^2_7\right],
\\ \nl &&b_3=-\frac{k}{3}\sinh^3\rho \sin\alpha dt\wedge d\alpha\wedge d\beta, \h k=const ,\ea where
$B$ is the relative radius of $AdS_4$ with respect to the seven-sphere. We choose to parameterize $S^7$ as in \cite{27}:
\ba\nl ds^2_7=4d\xi^2&+&\cos^2\xi\left(d\theta^2+d\phi^2+d\psi^2+2\cos\theta d\phi d\psi\right)\\ \nl &+& \sin^2\xi\left(d\theta^2_1+d\phi^2_1+d\psi^2_1+2\cos\theta_1 d\phi_1 d\psi_1\right).\ea
Now, consider the following membrane embedding of the type (\ref{sLA})
\ba\nl &&X^0(\tau,\delta,\s)\equiv t(\tau,\delta,\s)=\Lambda_0^0\tau,
\\ \nl &&X^1(\tau,\delta,\s)=Z^1(\s)=\rho(\s),
\\ \nl &&X^2(\tau,\delta,\s)=Z^2(\s)=\alpha(\s),
\\ \nl &&X^3(\tau,\delta,\s)\equiv \beta(\tau,\delta,\s)=\Lambda_0^3\tau + \La_1^3\delta+\La_2^3\s,
\\ \nl &&X^4(\tau,\delta,\s)\equiv \phi(\tau,\delta,\s)=\Lambda_0^4\tau+\La_1^4\delta+\La_2^4\s,
\\ \nl &&X^5(\tau,\delta,\s)\equiv \psi(\tau,\delta,\s)=\Lambda_0^5\tau+\La_1^5\delta+\La_2^5\s ,\ea
where in our notations $\mu=0, 3, 4, 5$, $a=1, 2$. 
The relevant background seen by the membrane is
\ba\nl ds^2 &=& l_{11}^2\left[-\cosh^2\rho dt^2 + d\rho^2 + \sinh^2\rho
\left(d\alpha^2 + \sin^2\alpha d\beta^2\right)\right.
\\ \nl &+&\left.B^2\left(d\phi^2+d\psi^2+2d\phi d\psi\right)\right]
\\ \nl &&b_{023}=-\frac{k}{3}\sinh^3\rho\sin\alpha.\ea
According to (\ref{if}), the fields induced on the membrane worldvolume are
\ba\nl &&G_{00}=-l_{11}^2\left[\left(\Lambda_0^0\right)^2\cosh^2\rho - \left(\Lambda_0^3\right)^2\sinh^2\rho\sin^2\alpha - B^2\left(\Lambda_0^4+\Lambda_0^5\right)^2\right],
\\ \nl &&G_{11}=l_{11}^2\left[\left(\Lambda_1^3\right)^2\sinh^2\rho\sin^2\alpha + B^2\left(\Lambda_1^4+\Lambda_1^5\right)^2\right],
\\ \nl &&G_{22}=l_{11}^2\left\{\rho'^2+\sinh^2\rho\left[\alpha'^2 + \left(\Lambda_2^3\right)^2\sin^2\alpha\right]+B^2\left(\Lambda_2^4+\Lambda_2^5\right)^2\right\},
\\ \nl &&G_{01}=l_{11}^2\left[\Lambda_0^3\Lambda_1^3\sinh^2\rho\sin^2\alpha + B^2\left(\Lambda_0^4+\Lambda_0^5\right)\left(\Lambda_1^4+\Lambda_1^5\right)\right],
\\ \nl &&G_{02}=l_{11}^2\left[\Lambda_0^3\Lambda_2^3\sinh^2\rho\sin^2\alpha + B^2\left(\Lambda_0^4+\Lambda_0^5\right)\left(\Lambda_2^4+\Lambda_2^5\right)\right],
\\ \nl &&G_{12}=l_{11}^2\left[\Lambda_1^3\Lambda_2^3\sinh^2\rho\sin^2\alpha + B^2\left(\Lambda_1^4+\Lambda_1^5\right)\left(\Lambda_2^4+\Lambda_2^5\right)\right],
\\ \nl &&B_{012}=\frac{k}{3}\Lambda_0^0\Lambda_1^3 \alpha'\sinh^3\rho\sin\alpha.\ea

For simplicity, we choose to work in diagonal worldvolume gauge $\lambda^i=0$ in which, in accordance with (\ref{rci}), we must have $G_{0i}=0$. There exist four types of solutions for these constraints, for the membrane embedding used:
\ba\label{Ia}  \Lambda_{i}^{3}=0,\h \Lambda_{i}^{5}= -\Lambda_{i}^{4},
\\ \label{Ib}  \Lambda_{i}^{3}=0,\h \Lambda_{0}^{5}= -\Lambda_{0}^{4},
\\ \nl \Lambda_{0}^{3}=0,\h \Lambda_{i}^{5}= -\Lambda_{i}^{4},
\\ \nl \Lambda_{0}^{3}=0,\h \Lambda_{0}^{5}= -\Lambda_{0}^{4}.\ea
Let us note that in the first two cases, which will be considered here, the induced $B$-field is zero, while for the last two, it is not. They will be considered in \cite{BR} along with other membrane configurations in diverse M-theory backgrounds.

Working in the framework of ansatz (\ref{Ia}), one obtains that $\det{G_{mn}}=\mathbf{G}=0$, i.e. this case corresponds to null membrane. The Nambu-Goto type action (\ref{NGa}) is identically zero, and our action simplifies to
\ba\nl S= \int d^{3}\xi\frac{G_{00}}{4\lambda^0}.\ea
The equations of motion (\ref{ema}) and the remaining constraint (\ref{00e}) take the form
\ba\nl \p_a G_{00}=0,\h G_{00}=0.\ea
Obviously, the constraint is the searched solution of the equations of motion.
Moreover, it gives the trajectory $\rho=\rho(\alpha)$:
\ba\nl \sinh\rho(\alpha)= \left[\frac{\left(B/\Lambda_0^0\right)^2\left(\Lambda_0^4+\Lambda_0^5\right)^2-1}
{1-\left(\Lambda_0^3/\Lambda_0^0\right)^2\sin^2\alpha}\right]^{1/2}.\ea
In the case under consideration, the constants of the motion (\ref{cm}) are zero, while the conserved momenta (\ref{cmom}) are connected with each other by the equality
\ba\nl \Lambda_0^0 E = \Lambda_0^3 S + \frac{2\lambda^0}{l_{11}^2B^2} J^2,\ea
where $E=-p_0$, $S=p_3$, $J=p_4=p_5$.

In the framework of ansatz (\ref{Ib}), one obtains that the Nambu-Goto type action (\ref{NGa}) is given by
\ba\nl &&S^{NG}= - T_2 l_{11}^3 B\left(\Lambda_{1}^{4}+\Lambda_{1}^{5}\right)\int d^{3}\xi \\ \nl &&\times
\sqrt{\left\{\sinh^2\rho\left[\left(\Lambda_0^0\right)^2 - \left(\Lambda_0^3\right)^2\sin^2\alpha\right] + \left(\Lambda_0^0\right)^2\right\}\left(\rho'^2 + \alpha'^2\sinh^2\rho\right)}.\ea
Our Lagrangian (\ref{old}) is
\ba\nl \mathcal{L}^{A}(\sigma) &=& -\frac{l_{11}^2}{4\lambda^0}\left\{\left(2\lambda^0T_2\right)^2 \left(l_{11}B\right)^2\left(\Lambda_{1}^{4}+\Lambda_{1}^{5}\right)^2\left(\rho'^2 + \alpha'^2\sinh^2\rho\right)\right.
\\ \nl &+& \left.\sinh^2\rho\left[\left(\Lambda_0^0\right)^2 - \left(\Lambda_0^3\right)^2\sin^2\alpha\right] + \left(\Lambda_0^0\right)^2\right\},\ea
which means that
\ba\nl &&K_{11} = -\left(2\lambda^0T_2\right)^2 \left(l_{11}^2 B\right)^2\left(\Lambda_{1}^{4}+\Lambda_{1}^{5}\right)^2\equiv -K^2,\h K_{22} = -K^2\sinh^2\rho,
\\ \nl &&A_a=0,\h V=l_{11}^2\left\{\sinh^2\rho\left[\left(\Lambda_0^0\right)^2 - \left(\Lambda_0^3\right)^2\sin^2\alpha\right] + \left(\Lambda_0^0\right)^2\right\}.\ea
The constants of the motion $\mathcal{P}^2_\mu$ are zero again. Now, the conserved quantities (\ref{cmom}) read
\ba\nl &&E=-p_0=\frac{l_{11}^2}{2\lambda^0}\Lambda_0^0 \int d^{2}\xi\cosh^2\rho,
\\ \nl &&S= p_3=\frac{l_{11}^2}{2\lambda^0}\Lambda_0^3 \int d^{2}\xi\sinh^2\rho\sin^2\alpha,
\\ \nl &&J= p_4=p_5=0.\ea

If we fix $\alpha$ or $\rho$, the membrane solution is given by (\ref{1dc}). Let us consider these particular cases.

Taking $\alpha=0$, we obtain
\ba\nl \rho(\sigma)= \ln\tan\left(\sigma/2A_{\rho}\right),\h
A_{\rho}=2\lambda^0T_2l_{11}B\frac{\Lambda_{1}^{4}+\Lambda_{1}^{5}}{\Lambda_{0}^{0}}.\ea
For $\alpha=\alpha_0\ne 0,\pi$, there exist two different solutions:
\ba\nl &&\rho(\sigma)= \frac{1}{2}\ln\frac{1+sn(\s/A_\rho)}{1-sn(\s/A_\rho)},
\h k^2 = 1-\left(\Lambda_0^3/\Lambda_0^0\right)^2\sin^2\alpha_0 \in(0,1);
\\ \nl && \tanh\rho(\sigma)=\frac{1}{\sqrt{1+k^2}}sn\left(\frac{\sqrt{1+k^2}}{A_\rho}\sigma\right),
\h k^2= \left(\Lambda_0^3/\Lambda_0^0\right)^2\sin^2\alpha_0 - 1 \in(0,1),\ea
where $sn(u)$ is one of the Jacobian elliptic functions.

Fixing $\rho=\rho_0\ne 0$, one obtains
\ba\nl &&\alpha(\s)= \arcsin\left[sn\left(\s/A_{\alpha}\right)\right],\h \alpha\in (-\pi/2,\pi/2),
\\ \nl &&A_{\alpha}=A_{\rho}\tanh\rho_0,\h k^2=\left(\Lambda_0^3/\Lambda_0^0\right)^2\tanh^2\rho_0\in (0,1).\ea

Now, let us turn to the general case, when none of the coordinates $\rho$ and $\alpha$ is kept fixed.
The equations of motion (\ref{ema}) and the yet unsolved constraint (\ref{00e}) may be written in the form
\ba\label{a1} &&\left(K_{11}\rho'\right)'-\frac{1}{2}\frac{\p K_{22}}{\p\rho}\alpha'^2 + \frac{1}{2}\frac{\p V}{\p \rho}=0,
\\ \label{a2} &&\left(K_{22}\alpha'\right)' + \frac{1}{2}\frac{\p V}{\p \alpha}=0,
\\ \label{00r} &&K_{11}\rho'^2 + K_{22}\alpha'^2 + V=0.\ea
(\ref{a1}) and (\ref{a2}), with the help of (\ref{00r}), can be transformed into
\ba\nl &&\left[\left(K_{11}\rho'\right)^2\right]' + \rho'\left[\frac{\p}{\p \rho}K_{11}V + \left(K_{22}\alpha'\right)^2\frac{\p}{\p \rho}\frac{K_{11}}{K_{22}}\right]=0,
\\ \nl &&\left[\left(K_{22}\alpha'\right)^2\right]' + \alpha'\frac{\p}{\p \alpha}K_{22}V=0,\ea
which possess first integrals of the type (\ref{fir}), (\ref{fia}).
In order to be able to give {\it explicit} solution, we set $\Lambda_0^3=0$ and obtain
\ba\nl \cosh\rho(\s)=\frac{A}{cn(C\s)},\ea
where $cn(u)$ is Jacobian elliptic function,
\ba\nl A=\sqrt{\frac{1}{2}\left[1+\sqrt{1+\left(\frac{2d}{l_{11}\Lambda_0^0 K}\right)^2}\right]},
\h C=\frac{l_{11}\Lambda_0^0}{K}\left[1+\left(\frac{2d}{l_{11}\Lambda_0^0 K}\right)^2\right]^{1/4},\ea
and $d$ is an arbitrary constant. The solution for the membrane trajectory is the following
\ba\nl \alpha(\rho)=\frac{d}{\left(A^2-1\right)CK^2} \left[A^2\Pi\left(\varphi, -\frac{1}{A^2-1}, k\right) - \left(A^2-1\right)F\left(\varphi, k\right)\right],\ea
where $\Pi$ and $F$ are elliptic integrals, and
\ba\nl \varphi=\arccos\left(\frac{A}{\cosh\rho}\right),\h k=\left[1+\left(\frac{2d}{l_{11}\Lambda_0^0 K}\right)^2\right]^{-1/4} \sqrt{\frac{1}{2}\left[\sqrt{1+\left(\frac{2d}{l_{11}\Lambda_0^0 K}\right)^2}-1\right]}.\ea

The condition $\Lambda_0^3=0$ leads to $S=0$, and the membrane energy $E$ remains the only nontrivial conserved quantity. On the obtained solution, it is given by
\ba\nl E= A^2E_0 + \frac{\pi CK^2}{2\lambda^0\Lambda_0^0}
\left[\frac{sn(2\pi C)dn(2\pi C)}{cn(2\pi C)} - E(k)\right],\h E_0=E_{\rho=0},\ea
where $E(k)$ is the complete elliptic integral of second kind.

\vspace*{.5cm} {\bf Acknowledgments} \vspace*{.2cm}

This work is supported by NSFB grant under contract $\Phi1412/04$.




\end{document}